\documentclass[prl,aps,showpacs,superscriptaddress,amsmath,twocolumn,floatfix]{revtex4}
\usepackage{graphicx}

\begin{document}

\title{Van der Waals Density Functional for Layered Structures}

\author{H.~Rydberg}%,$^{1}$
\affiliation{
Department of Applied Physics, Chalmers University of Technology and G{\"o}teborg University, 
SE-412 96 G\"{o}teborg, Sweden}
  \author{M.~Dion}%$^{2}$
\affiliation{Center for Materials Theory, 
Department of Physics and Astronomy, Rutgers University, Piscataway, New Jersey 08854-8019}
  \author{N.~Jacobson}%,$^{1}$  
\affiliation{
Department of Applied Physics, Chalmers University of Technology and G{\"o}teborg University, 
SE-412 96 G\"{o}teborg, Sweden}
  \author{E.~Schr\"oder}%,$^{1}$
\affiliation{
Department of Applied Physics, Chalmers University of Technology and G{\"o}teborg University, 
SE-412 96 G\"{o}teborg, Sweden}
  \author{P.~Hyldgaard}%,$^{1}$
\affiliation{
Department of Applied Physics, Chalmers University of Technology and G{\"o}teborg University, 
SE-412 96 G\"{o}teborg, Sweden}
  \author{S.I.~Simak}%,$^{1}$
\affiliation{
Department of Applied Physics, Chalmers University of Technology and G{\"o}teborg University, 
SE-412 96 G\"{o}teborg, Sweden}
  \author{D.C.~Langreth}%,$^{2}$ 
\affiliation{Center for Materials Theory, 
Department of Physics and Astronomy, Rutgers University, Piscataway, New Jersey 08854-8019}
  \author{B.I.~Lundqvist}\thanks{To whom 
  correspondence should be addressed; 
E-mail: lundqvist@fy.chalmers.se.}%$^{1}
\affiliation{
Department of Applied Physics, Chalmers University of Technology and G{\"o}teborg University, 
SE-412 96 G\"{o}teborg, Sweden}

\date{August 6, 2003}

%%%%%%%%%%%

\begin{abstract}
To understand sparse systems we must account for both strong local atom bonds and 
weak nonlocal van der Waals forces between atoms separated by empty space. A fully 
nonlocal functional form [H. Rydberg, B.I. Lundqvist, D.C. Langreth, and M. Dion, 
Phys.\ Rev.~B {\bf 62}, 6997 (2000)] of density-functional theory 
(DFT) is applied here to the layered systems graphite, boron nitride, and molybdenum 
sulfide to compute bond lengths, binding energies, and compressibilities. These 
key examples show that the DFT with the generalized-gradient approximation does not 
apply for calculating properties of sparse matter, while use of the fully nonlocal 
version appears to be one way to proceed.  
\end{abstract}


\maketitle

Calculations of structure and other properties of 
sparse systems must account for {\it both} strong local atom bonds {\it and} 
weak nonlocal van der Waals (vdW) forces between atoms separated by empty space. 
Present methods are unable to describe the true interactions 
of sparse systems, abundant among materials and molecules.
Key systems, like  graphite, BN, and MoS$_2$, have layered structures. While 
today's standard tool, density-functional theory (DFT), has broad application, 
the common local (LDA) and semilocal density approximations (GGA) 
\cite{PW91,PBE,ZhYa98,RPBE} for exchange and correlation, $E_\text{xc}[n]$, 
fail to describe the interactions at sparse electron densities. Here we show 
that the recently proposed density functional~\cite{RyLuLaDi00} with nonlocal 
correlations, $E^\text{nl}_\text{c}[n]$, gives separations, binding energies, 
and compressibilities of these layered systems in fair agreement with 
experiment. This planar case bears on the development of 
vdW density functionals for general geometries~\cite{RyThesis,DiRyLuLa03}, 
as do asymptotic vdW functionals \cite{everything}.

Figure 1 with its `inner surfaces' defines the problem: voids of 
ultra-low density, across which electrodynamics leads to vdW coupling. 
This coupling depends on the polarization properties of the layers themselves, 
and {\it not} on small regions of density overlap between the layers, 
excluding proper account in LDA or GGA. For large interplanar separation $d$ 
the vdW interaction energy between planes behaves as $-c_4/d^4$, while LDA or 
GGA necessarily predicts an exponential falloff. Layers rolled up to form two 
(i) nanotubes with parallel axes a distance $l$ apart interact as $-c_5/l^5$, 
or (ii) balls ({\it e.g.}, C$_{60}$), a distance $r$ apart, as $-c_6/r^6$. 
If by fluke an LDA or GGA were to give the correct equilibrium for one 
shape, it would necessarily fail for the others. The simple expedient 
of adding the standard asymptotic vdW energies as corrections to the 
correlation energy of LDA or GGA also fails. The true vdW interaction 
between two close sheets must be (i) substantially stronger (Fig.~1), 
(ii) {\it seamless}, and (iii) saturate as $d$ shrinks (Fig.~2).

Like earlier work directly calculating nonlocal correlations between two 
jellium slabs~\cite{DoWa99}, the vdW density functional (vdW-DF) 
theory~\cite{RyLuLaDi00} used here exploits assumed planar symmetry. It 
divides the correlation energy functional into two pieces, $E_\text{c}[n] = 
E_\text{c}^0[n]+E_\text{c}^\text{nl}[n]$, where $E_\text{c}^\text{nl}[n]$ 
is defined to include the longest ranged or most nonlocal terms that give 
the vdW interaction and to approach zero in the limit of a slowly varying 
density. The term $E_{\rm c}^0[n]$ is also nonlocal, but approaches the LDA 
in this limit. The two terms are approximated differently. Here the LDA is 
used for $ E_\text{c}^0[n]$, as the LDA should be much more accurate with 
the principal longest range terms separated off. Ultimately we plan to 
derive a gradient or GGA expansion appropriate to $ E_\text{c}^0[n]$, but 
the resulting corrections are expected to be small. Thus we use 
\begin{equation}
E_\text{c}[n]\approx E_\text{c}^\text{LDA}[n]+E_\text{c}^\text{nl}[n].
\label{eq:ec1}
\end{equation}
Long range terms are less sensitive to details of the system's 
dielectric response. Thus very simple approximations for the dielectric 
function are made for $ E_\text{c}^\text{nl}[n]$. We care to make 
the polarization properties of a single layer come out reasonably.

For systems with planar symmetry, the predominant component of the 
nonlocal correlation energy $E^\text{nl}_\text{c}$ giving the vdW forces can 
be determined simply by comparing the solutions of the Poisson equation 
$\nabla\cdot(\epsilon\nabla\Phi) = 0$ and the Laplace equation 
$\nabla^2 \Phi = 0$~\cite{RyLuLaDi00}. The crux of the approximation is 
the use of a simple plasmon-pole model for the dielectric function,
\begin{equation}
\label{eq:epsilon}
\epsilon_k(z,iu) = 
1+\frac{\omega_{\rm p}^2(z)}{u^2+(v_{\rm F}(z) q_k)^2/3+q^4_k/4},
\end{equation}
where $\omega_{\rm p}^2(z) = 4 \pi n(z)e^2/m$, and 
$mv_{\rm F}(z)=(3 \pi^2 n(z))^{1/3}$ are functions of the local density $n(z)$. 
The 2D wave vector perpendicular to the $z$ direction is $\mathbf{k}$, $iu$ 
is the imaginary frequency and $q_k^2=k^2 + q_\perp^2$ mimicks the 3D wave 
vector, with the $z$ direction accounted for by the $n(z)$ dependence and 
$q_\perp$ taken as a constant.

%%%%%%%%%
\begin{figure}
\centerline{\includegraphics[width=.4\textwidth]{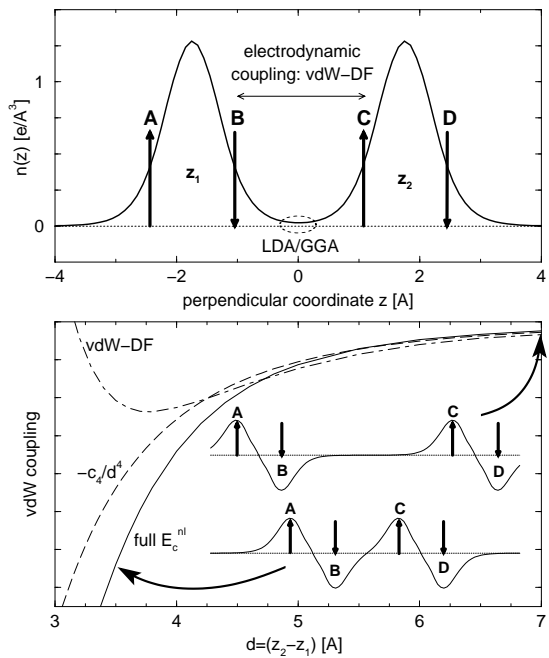}}
\caption{\label{fig1}
Schematics of the vdW forces in sparse, layered materials. 
{\it Top panel\/}: 
Laterally averaged electron-density profile of two graphene layers at 
equilibrium separation $d\approx$ 3.5 {\AA}. The dotted ellipse indicates the 
region determining the interlayer interaction within LDA or GGA. It is 
unable to describe the true long range interaction (horizontal arrow) 
between correlated charge fluctuations (extrema illustrated by those of the 
finite-frequency charge response, and labelled by 
vertical arrows). {\it Bottom panel}: Full $E_\text{c}^\text{nl}$  (solid curve) vs.\ 
the asymptotic form  of $E_\text{c}^\text{nl}\rightarrow-c_4/d^4$ (dashed line).
Also shown: full ground state interaction energy (dash-dotted line). 
{\it Insets}:  Sketches of the correlated charge fluctuations for different layer
separations, explaining why $ |E_\text{c}^\text{nl}| >c _4/d^4$. In the asymptotic 
region (top inset), attractive interactions BC and AD are nearly cancelled by the 
repulsive interactions AC and BD, leaving little more than the weak dipolar 
interaction.  Near equilibrium separation (bottom inset), the repulsive 
interactions AC and BD remain specified by the layer separation $d$, but the 
relative strengthening of attractive interaction BC %(distance $\overline{BC} < d$) 
substantially overcompensates the relative weakening of the attractive 
interaction  AD. At still smaller $d$, $E_\text{c}^\text{nl}$ becomes weaker and 
saturates (Fig. 2, inset).
}
\end{figure}

%%%

Our scheme, applied to jellium surfaces in Ref.~\onlinecite{RyLuLaDi00}, is 
used here for layered systems, for simplicity a two-layered system 
({\it e.g.,} two parallel graphene sheets) arranged perpendicular to the 
$z$ axis, at positions $z=0$ and $z=d$, respectively. The first step is to 
calculate the density in DFT with a suitable approximation, which we take 
to be an appropriate flavor of the GGA, and to average it in the lateral 
direction to yield $n(z)$. The next two steps represent the key to the 
success of our approximation. First one calculates from first principles 
the perpendicular static polarizability $\alpha$ of a {\it single} layer 
of the material (a graphene layer in this example) in the ground-state DFT 
scheme with the above GGA flavor. Then one fixes the constant $q_\perp$ so 
that dielectric model (\ref{eq:epsilon}) gives precisely the same 
polarizability, that is 
\begin{equation}
 \alpha = \frac{1}{4\pi} \int \! dz \left[1-\frac{1}{\epsilon_0(z,0)}\right].
 \label{alpha}
\end{equation}
This step mitigates the errors introduced both by the lateral averaging and 
the dielectric model. The leading vdW interaction is proportional to an 
integral over the square of the dynamic polarizabilities. Previous work in 
the asymptotic limit \cite{everything} shows that simple, properly scaled 
dielectric functions giving the correct static polarizabilities form good 
approximations to the dynamic ones, hence giving accurate vdW interactions.

%%%%%%%%%%%%%%%%%%%%%%%%%%%%%%%%%

\begin{figure}
\centerline{\includegraphics[width=.4\textwidth]{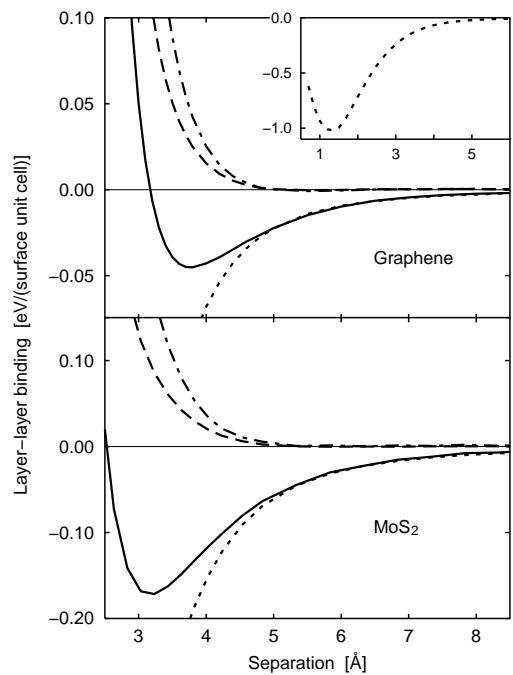}}
\caption{\label{fig2}
Calculated vdW-DF results for the interlayer binding (adhesion) 
between two sheets of graphene ({\it upper panel}) and MoS$_2$ 
({\it lower panel}) as functions of the interlayer separation $d$. In each 
panel it is compared (solid curves) with the underlying GGA (dashed curves),  
and with the adhesion {\it in the absence of} the vdW-DF term $E_\text{c}^\text{nl}$
(dash-dotted curves) calculated using only the first two terms of Eq.~(\ref{eq:exc1}) 
for $E_\text{xc}$. The nonlocal-correlation vdW contribution $E_{c}^\text{nl}$ alone 
is shown separately (dotted curves). The figure shows that no binding results 
from exchange or local correlation; the vdW interactions provides adhesion in 
fair agreement with experiment (Table I). The {\it inset} shows that 
our calculated value of $E_{c}^\text{nl}$  
saturates for small $d$ values.
}
\end{figure}

%%%%%%%%%%%%%%%%%%%%%%%%%%%%%%%%%

The correlation energy can be calculated from the charge response%
~\cite{couplingconstantintegration}. With planar symmetry only the 
response to an arbitrarily charged sheet $\rho \propto\exp(i\mathbf{k} \cdot 
\mathbf{r})\delta (z-z_-)$ placed at an offset $z_-$ ($z_- \ll 0$) from one 
end of the sample needs to be calculated. Integrating Poisson's equation 
with $\epsilon$ as in (\ref{eq:epsilon}) one finds the $z$ component of the 
electric field  $E_z(z_+)$ at $z=z_+\gg d$.  Then $E_{\rm c}^\text{nl}$ is given 
by \cite{RyLuLaDi00}
\begin{equation}
\label{eq:enl}
E_c^\text{nl} = - A
\int_0^\infty \! \frac{du}{2\pi} \,
\int\! \frac{d^2k}{(2\pi)^2}
\ln
\frac{E_z(z_+)}{E^0_z(z_+)},
\end{equation}
where $A$ is the lateral area and $E^0_z(z_+)$ is the electric field 
component from the charge sheet at $z=z_-$ in the absence of the sample. In 
Eq.~(\ref{eq:enl}) the spatial dependences of the two $E$'s cancel in the 
ratio, leaving only the $k$ dependence. For each of the bulk solids considered 
here we apply this method to 32 and 30 layer slabs to give, by subtraction, 
a well converged per layer value of $E_{\rm c}^{\rm nl}$.

%%%%%%%%%%%%%%%%%%%%%%%%%%%%%%%%%

\begin{figure}
\includegraphics[width=.4\textwidth]{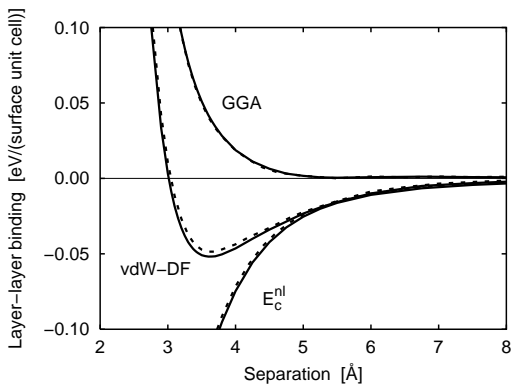}
\caption{\label{fig3}
Comparison of the vdW-DF layer-to-layer binding in the staggered 
bulk hexagonal-BN system (solid lines) and adhesion between two 
sheets of BN (broken lines) shown as functions of the interlayer 
separation $d = c/2$. In both descriptions the GGA gives 
unphysical results for bond and equilibrium separation. 
The total 
adhesion is dominated by the contribution of the vdW interaction 
$E_\text{c}^{\text nl}$. The nearest-neighbor interlayer binding 
(broken lines) is seen to strongly dominate the bulk adhesion 
(solid lines) for layered materials.  
}
\end{figure}

%%%%%%%%%%%%%%%%%%%%%%%%%%%%%%%%%

The approximation for the exchange energy functional $E_{\rm x}[n]$ 
should be consistent with our approximation for $E_{\rm c}^0[n]$, 
that is a local or semilocal one. We use a version that best 
approximates exact density functional exchange,  
namely Zhang and Yang's~\cite{ZhYa98} (ZY) ``revPBE'' exchange functional 
$E_\text{x}^\text{ZY}[n]$, where
the parameter controlling 
the large gradient limit in the PBE exchange functional~\cite{PBE} is fitted 
to exact density-functional exchange calculations. Unlike exchange 
\cite{ScolesPlus} in PW91 \cite{PW91}, PBE \cite{PBE}, and RPBE~\cite{RPBE}, 
$E_\text{x}^\text{ZY}$ shows no tendency to bind any of the vdW systems we 
have tried it on, unless the vdW correlation is actually included in the 
calculation. We thus take 
\begin{equation}
E_{\rm xc}[n] = E_\text{x}^\text{ZY}[n]+ E_\text{c}^\text{LDA}[n]+E_\text{c}^\text{nl}[n]. 
\label{eq:exc1}
\end{equation}
The first two terms form an important and recommended starting point 
for testing nonlocal vdW functionals. They assure a vdW attraction that 
actually comes from terms that in principle are capable of giving it.

%%%%%%%%%%%%%%%%%%%%%%%%%%%%%%%%%

\begin{table}
\caption{\label{table1}
Properties of graphite, 
BN, and MoS$_2$, calculated with vdW-DF and compared to experimental data 
in parentheses, when available.  The table shows the geometry ($a,c$), 
binding energy ($E_0$), bulk modulus ($B_0$), and elastic constant ($C_{33}$). 
GGA (not shown) binds very weakly at unphysically large $c$ or not at all, 
depending on substance and GGA flavor.}
\begin{ruledtabular}
\begin{tabular}{lccc}
& Graphite& BN& MoS$_2$ \\
\hline
$a$ [\AA]& 2.47 (2.46)\footnotemark[1] & 2.51 (2.50)\footnotemark[2] & 3.23 (3.160)\footnotemark[3]\\ 
$c$ [\AA]& 7.52 (6.70)\footnotemark[4]& 7.26 (6.66)\footnotemark[2] & 12.6 (12.29)\footnotemark[3] \\
$E_0$ [meV/at]& 24 ($35 \pm 10$)\footnotemark[5] & 26& 60\\
$B_0$ [GPa]& 12 ($\sim$$33$)\footnotemark[1] & 11 & 39\\
$C_{33}$ [GPa]& 13 (37-41)\footnotemark[1] & 11 & 49\\
\end{tabular}
\end{ruledtabular}
\footnotetext[1]{Ref.~\onlinecite{Bornstein80}.}
\footnotetext[2]{Ref.~\onlinecite{KeKrHa99}.}
\footnotetext[3]{Ref.~\onlinecite{MoS2exp}.}
\footnotetext[4]{Ref.~\onlinecite{BaMa55}.}
\footnotetext[5]{Ref.~\onlinecite{BCCZLC98}.}
\end{table}
%%%%%%%%%%%%%%%%%%%%%%%%%%%%%%%%%

The GGA is used in order to have a good account of strong valence bonds and 
densities of individual layers, and the above scheme is implemented by 
self-consistently calculating the energy $E^\text{GGA}$ and density 
$n^\text{GGA}$, typically in the revPBE flavor 
of GGA. The total energy, as a function of the lattice parameters, is then 
calculated as $E\approx E^{\rm GGA}[n^{\rm GGA}] +\Delta[n^{\rm GGA}]$, 
where $\Delta[n]= E_{\rm c}^{\rm nl}[n]  - E_{\rm c}^{\rm grad}[n]$ and 
$ E_{\rm c}^\text{grad}[n] = E_{\rm c}^\text{GGA}[n] - E_{\rm c}^\text{LDA}[n]$. 
This approximation treats $\Delta$ as a post GGA perturbation. The GGA calculations 
are done using the plane-wave pseudopotential~\cite{corecomment} DFT code 
dacapo~\cite{DACAPO}. 

The three materials considered here have layered structures 
with a soft direction perpendicular to the layers 
\cite{separation}. Strong covalent bonds occur between the atoms within 
the hexagonal layers (lattice constant $a$), while weak vdW interactions 
occur between the layers. Graphite has a staggered or A-B type stacking
with the intralayer C atoms in regular, planar hexagons stacked with 
corners on top of centers. Boron nitride, isoelectronic with graphite, 
has no shift in hexagon locations but B and N atoms in alternate positions 
along the $c$-direction. A similar alternation occurs between  Mo and S 
atoms in the A-B type structure of MoS$_2$ \cite{MoS2 footnote}. 

Calculations are also performed for bilayers of the three systems. The 
binding-energy curve of two parallel graphene sheets~\cite{DDSSG88}, 
{\it i.e.} the total-energy difference at separation $d$ and at infinite 
separation, respectively, for varying $d$ (Fig.~2) gives in the 
approximation~\cite{RyLuLaDi00} with vdW-DF, Eq.~(\ref{eq:exc1}), a close 
relation to experimental findings for binding energy and equilibrium 
distance. The GGA curve, on the other hand, is completely wrong, which 
we blame on absence of vdW effects.  The inset shows that our 
$E_c^\text{nl}$ contribution gives stronger binding than the 
traditional asymptotic vdW interaction. The binding-energy curve for 
two parallel BN sheets (Fig.~3) differs little from that of bulk BN 
as a function of lattice constant $c$. At equilibrium the calculated 
$E_\text{c}^\text{nl}$ difference  is $\sim 4$\%, attributable mostly to  
2nd nearest layers in the bulk material \cite{fourpercent}.

Bulk-modulus values are computed together with other structure ($a$,$c$) 
and bonding properties \cite{subtraction} (Table I). It is crucial to 
densely sample the region of ($a$,$c$) values around the optimal structure 
($a_0$, $c_0$), and we use a new method \cite{ZiSc03} for direct evaluation 
of both structure, binding energy, and bulk modulus $B_0$ in the relevant 
range of $a$ and $c$ values \cite{B0comment}. GGA values were also 
computed but not shown. For all three materials the various GGA flavors 
give no binding or bind very weakly at unreasonably large separation. 
The vdW-DF, on the other hand, gives values for lattice parameters and 
cohesive energy in good agreement and for bulk modulus in fair agreement 
with experimental values, when available.

In conclusion, we recommend the replacement of GGA as a standard 
method in total-energy calculations with vdW-DF as given in 
Eq.~(\ref{eq:exc1}) for descriptions of layered systems that contain 
sparse electron distributions. This will give the right qualitative 
character of soft bonds, including saturation of the vdW potential 
at small separations, and even quantitative ones, like physical 
values of bond lengths, binding energies, and compressibilities 
\cite{improvements}.

We are grateful to K.~W.~Jacobsen for suggesting tests on MoS$_2$ 
layers \cite{Boll}. Financial support from the Swedish Foundation for 
Strategic Research via Materials Consortia \#9 and ATOMICS, STINT, and 
the Swedish Scientific Council are gratefully acknowledged. Work by 
M.D.~and D.C.L.~supported in part by NSF Grant DMR 00--93070.

%%%%


\begin{thebibliography}{99}
\bibitem{couplingconstantintegration}
The coupling constant integration in $E^{\rm nl}$ is performed 
via the following approximation: 
$\epsilon -1 \approx \lambda [(\epsilon - 1)/\lambda]_{\lambda 
=1}$. This approximation gives the long range van der Waals 
interaction exactly \cite{DiRyLuLa03} for dielectric functions
of the type used here.  The $\lambda$ dependence comes
via  $q_\perp$.
\bibitem{corecomment}
Only the valence density was used in the functional, because the lateral
averaging unphysically lowers the core density and increases the size
of the core, yielding an unphysically large vdW contribution.  Nevertheless,
tests with the full density showed negligible effect on the geometry,
and at most a 30\% increase in well depth.
\bibitem{separation}
Separation $d=c/2$, except for MoS$_2$, where  $d$ is defined as 
the distance from the perpendicular   
position of the top-most sulfur atom of one layer to the position of 
the bottom-most sulfur atom of the next.
\bibitem {MoS2 footnote}
In the MoS$_2$ interlayer calculations, each MoS$_2$ layer 
is treated as a composite material  (the  Mo-S perpendicular separation 
is  $\sim$1.58 {\AA}). A composite $q_{\perp}$ value ($\sim$$0.58$),   
 identical in both MoS$_2$ layers, is obtained  by applying an 
external field across a single MoS$_2$ composite and using GGA to 
calculate  the static polarization.
 In the interaction study 
the vdW-DF energy is then calculated as a function of the separation 
between the two MoS$_2$-layer composite materials, the atom positions 
within each of the MoS$_2$-layer composites being kept fixed.
\bibitem{fourpercent}
The 4\% figure may be estimated as $1/(1.5\times 2^4)$, with the divisor of 1.5
occurring because the 50\% enhancement (Fig.~1) is not applicable at the 2nd
layer distance.
\bibitem{B0comment}
The value of 33 GPa reported  
earlier~\protect\cite{SurfSci03} was incorrect.
\bibitem{improvements}
Future work that may remove the remaining differences  includes
(i) time dependent DFT and other calculations of the positions 
of the polarization peaks (Fig.~\ref{fig1}) followed by 
improvements to the approximation Eq.~(\ref{eq:epsilon}), 
(ii) derivation of the gradient corrections appropriate to 
$E_c^0$  with corresponding modifications to Eq.~(\ref{eq:ec1}), and 
(iii)  implementation of exact exchange with corresponding
modifications to Eq.~(\ref{eq:exc1}).
\end{thebibliography}
\end{document}